\documentclass[acmlarge]{acmart}

\usepackage{multirow}
\usepackage{graphicx}
\usepackage{amsmath}
\usepackage{graphics}
\usepackage{array}
\usepackage{multirow}
\usepackage{enumitem}
\usepackage{url}
\usepackage{xcolor,colortbl}
\usepackage{hhline}
\usepackage{threeparttable}
\usepackage{tabularx}
\usepackage{caption}
\usepackage{rotating}
\AtBeginDocument{%
  \providecommand\BibTeX{{%
    \normalfont B\kern-0.5em{\scshape i\kern-0.25em b}\kern-0.8em\TeX}}}

\setcopyright{acmcopyright}
\copyrightyear{2023}
\acmYear{2023}
\acmDOI{10.1145/1122445.1122456}

\acmMonth{5}

\begin{document}

\title{A Comprehensive Picture of Factors Affecting User Willingness to Use Mobile Health Applications}

\author{Shaojing Fan}
\affiliation{%
	\institution{National University of Singapore}
	\country{Singapore}
}
\email{dcsfs@nus.edu.sg}

\author{Ramesh C. Jain}
\affiliation{%
	\institution{University of California}
	\country{USA}
}
\email{jain@ics.uci.edu}

\author{Mohan S. Kankanhalli}
\affiliation{%
	\institution{National University of Singapore}
	\country{Singapore}
}
\email{mohan@comp.nus.edu.sg }

\renewcommand{\shortauthors}{Fan and Jain, et al.}

\begin{abstract}
	
Mobile health (mHealth) applications have become increasingly valuable in preventive healthcare and in reducing the burden on healthcare organizations. The aim of this paper is to investigate the factors that influence user acceptance of mHealth apps and identify the underlying structure that shapes users' behavioral intention. An online study that employed factorial survey design with vignettes was conducted, and a total of 1,669 participants from eight countries across four continents were included in the study. Structural equation modeling was employed to quantitatively assess how various factors collectively contribute to users' willingness to use mHealth apps. The results indicate that users' digital literacy has the strongest impact on their willingness to use them, followed by their online habit of sharing personal information. Users' concerns about personal privacy only had a weak impact. Furthermore, users' demographic background, such as their country of residence, age, ethnicity, and education, has a significant moderating effect. Our findings have implications for app designers, healthcare practitioners, and policymakers. Efforts are needed to regulate data collection and sharing and promote digital literacy among the general population to facilitate the widespread adoption of mHealth apps.

\end{abstract}



\maketitle

\makeatletter
\def\hlinew#1{%
	\noalign{\ifnum0=`}\fi\hrule \@height #1 \futurelet
	\reserved@a\@xhline}


\section{Introduction}

The increasing prevalence of smartphones and wearable devices has created a growing demand for mobile applications. Mobile health (mHealth) applications, in particular, have become increasingly popular in response to both the recent Covid-19 pandemic and the aging global population. Previous studies have demonstrated the usefulness of mHealth apps, which have become a crucial avenue for healthcare service providers to deliver healthcare services \cite{heerden2012point}. Additionally, these applications offer benefits in promoting education and awareness of healthy lifestyles \cite{anderson2016mobile}.

Despite the usefulness of mHealth apps, studies have demonstrated that people exhibit varying levels of willingness and openness to use them. For instance, concerns have been raised by gynecologists about the security of mobile phone applications and the protection of patients' personal information \cite{gong2021surgical}. Furthermore, Alazzam and colleagues discovered that younger physicians were more likely to incorporate digital healthcare technology into their practice than their older counterparts \cite{alazzam2021survey}. Other studies have suggested that factors such as age and the number of follow-up visits were negatively correlated with patients' willingness to utilize mHealth applications \cite{dai2017willingness}.

This study aims to investigate the willingness of the general population to use mHealth apps, as well as the factors that influence such willingness. To achieve this, we conducted online surveys between December 2021 and March 2022 in eight countries across four continents: the United States (US), United Kingdom (UK), Germany, India, China, Singapore, New Zealand, and Australia. Our research was based on a vignette design of hypothetical mHealth apps, inspired by the authors' previously developed app \cite{oh2017multimedia,Personicleap}, as well as other existing apps from around the world. Specifically, we presented each participant with eight versions of an mHealth app, each with varying features in seven areas: app functionality, type of rewards for using the app, type of data collected by the app, where the collected data is stored, with whom the collected data is shared, the app's privacy protection measures, and user control of app data management. Participants indicated the likelihood of using each version of the app on their smartphone, and provided reasons for their decision. At the end of the experiment, participants provided information about their daily phone usage, online behavior, concern about privacy, and prior experience with mHealth apps (see Appendix C for the complete questionnaire). Through a factorial survey design with vignettes, we were able to simultaneously test the effect of several app characteristics on users' acceptance.

\section{Related Work}

In this section, we provide an overview of the benefits and challenges of mHealth apps in the market, as well as a review of prior research on the factors that influence user willingness to use mHealth apps.

\subsection{Benefits and challenges of mHealth apps}

mHealth apps are being used in a variety of ways, such as disease prevention, health promotion, patient education, self-management, and remote monitoring \cite{kao2017consumer}. Numerous studies have reported various benefits of mHealth apps, such as increased access to healthcare \cite{qudah2019influence,vaitkiene2022digital}, improved patient engagement \cite{singh2016developing}, and more efficient healthcare delivery \cite{eng2013promise,karatacs2022effectiveness}. Additionally, mHealth apps allow individuals to monitor their own health, track their progress, and receive personalized feedback and recommendations \cite{anderson2016mobile}.

While mHealth apps offer many potential benefits, they also face several challenges. Studies by Baig et al. \cite{baig2015mobile} and Newaz et al. \cite{newaz2021survey} have identified security and privacy issues as major challenges facing mHealth apps. Jaime and colleagues \cite{benjumea2020privacy} conducted a privacy assessment of mHealth apps based on 24 selected articles published from 2014 onwards. They found that despite great progress made in raising awareness of privacy and security in mHealth apps, there is still much to be done. Two major barriers identified were the diversity of mHealth apps and the lack of standard evaluation criteria. In addition, Tangari and colleagues \cite{tangari2021mobile} analyzed 20,991 mHealth apps on the Google Play Store and found serious problems with privacy and inconsistent privacy practices in mHealth apps. They urged clinicians to be aware of these issues and to articulate them to patients when determining the benefits and risks of using these apps.

\subsection{Factors Influencing User Acceptance of mHealth apps} \label{sec:relatedfactors}

While mHealth apps have great potential, some people may be hesitant to use them due to various reasons. These include concerns about the privacy and security of personal information \cite{zahra2018factor,aljedaani2021challenges}, a lack of trust in the accuracy of the information provided by the app, and concerns about the cost of the app \cite{woldaregay2018motivational}.

Previous studies have indicated that the willingness to use mHealth apps can be affected by various factors, such as age, gender, health status, and technological ability \cite{virella2019mobile,nunes2019individual}. However, the results from these studies are not entirely conclusive. One study by Andreia et al. \cite{nunes2019individual} found that older users exhibited a higher level of conscientiousness and behavioral intention to use mHealth apps. On the other hand, Virella et al. \cite{virella2019mobile} reported that younger individuals were more likely to use mHealth apps than older adults. Gender also plays a role in the relationship between personality traits and the behavioral intention to use mHealth apps, according to Nunes et al. \cite{nunes2019individual}. While Zhang et al. \cite{zhang2014understanding} found that males had a higher level of intention to adopt mHealth compared to females, a later study by Bol et al. \cite{bol2018differences} reported no gender effect on aggregated mHealth app use. Furthermore, Ernsting et al. \cite{ernsting2017using} found that individuals with chronic conditions were more likely to use mHealth apps. However, the study by Robbins et al. \cite{robbins2017health} suggested that individuals with mHealth apps were not more likely to have chronic health conditions compared to those without.

The previous studies have generated inconsistent results. In our research, we expand on the previous research by examining a more extensive range of factors that affect the acceptance of mHealth apps, including those that have not been well researched, such as users' cultural and educational backgrounds, as well as the data processing and storage of mHealth apps. Furthermore, we examine how users' prior experience with Covid-19 affects their willingness to use mHealth apps.

\section{Methodology} \label{sec:method}

\subsection{Research Question and Hypotheses}   \label{sec:hypotheses}

In this subsection, we describe our research questions, and propose hypotheses under each research question.\\

\textbf{RQ1:} What are the factors that influence users' willingness to use mHealth apps?\\

In this study, our goal is to present a comprehensive and unified perspective on the potential factors that impact user willingness to use mHealth apps. We propose that four broad categories of variables will influence this willingness. The first category is demographic background, including age, gender, and educational background. Secondly, user intention is influenced by perceived benefits and interest in the mHealth app, while perceived costs decrease willingness. Additionally, since mHealth apps handle users' health data, privacy concerns and the risk of disclosure become more prominent. Therefore, trust in the data collection and sharing process and privacy concerns will be of significant importance to users. Finally, the willingness to use mHealth apps will be impacted by factors related to users' technological background. These factors include user experience and comfort with mobile technology and feelings of control over the data management. We describe four hypotheses based on these factors below.

Previous studies have shown inconsistent results regarding the influence of age, gender, and health status on user willingness to use mHealth apps. While some studies have reported significant effects of these factors \cite{zhang2014understanding,ernsting2017using,nunes2019individual,virella2019mobile}, a few others have indicated limited effects \cite{robbins2017health,bol2018differences} (refer to Sec \ref{sec:relatedfactors} for a more detailed discussion). Recently, Utz and colleagues \cite{utz2021apps} reported that countries of residence also impact user preferences for the collection of personalized data or anonymity, with Chinese participants preferring the former and German and US participants favoring the latter. The variation in the impact of demographic background across studies may be partly due to different experimental settings and contexts. In our study, we hypothesize that if a demographic factor influences user views and preferences for mHealth apps, it is likely to impact user willingness to use mHealth apps. Based on this, we propose our first hypothesis,\\

\textbf{H1:} The willingness of users to use mHealth apps is influenced by their demographic background, including factors such as age, gender, educational background, and health status. \\

Previous research has shown the importance of perceived benefits and interest in user willingness of using mHealth apps \cite{boudreaux2014evaluating,wu2022factors}. Therefore, our second hypothesis is, \\

\textbf{H2:} Users are more willing to use an mHealth app if they find it beneficial to their health, or if they can get financial rewards for using the app. \\

Privacy and security are paramount for any mobile app, particularly mHealth apps that deal with sensitive and personal information such as users' health history, medication usage, and location data \cite{wang2021influencing}. Prioritizing privacy and security in an mHealth app can increase user willingness to use it. A study by Zhou and colleagues \cite{zhou2019barriers} revealed that most study participants had concerns about their privacy when using mHealth apps and expressed their preferences for security features, such as regular password updates, remote wipe, user consent, and access control. Based on these findings, we propose the third hypothesis in this study: \\

\textbf{H3:} Users' willingness to use an mHealth app is positively correlated with their trust in the app's privacy and security measures.\\

The technological background of a user can refer to their experience with technology and their level of comfort in using it. Previous studies have suggested a positive correlation between users' technological background and their willingness to use mHealth apps \cite{leigh2020barriers}. For instance, a study by Jaana and colleagues \cite{jaana2020comparison} revealed that older adults were less likely to use mobile health apps due to a lack of familiarity with technology. Thus, we propose the fourth hypothesis: \\

\textbf{H4:} Greater familiarity with technology is positively associated with user willingness to use mHealth apps. \\

\textbf{RQ2:} How do various factors jointly contribute to users' intention to use mHealth apps?\\

Previous research suggests that multiple factors interact with each other in influencing users' intention to use mHealth apps. For instance, gender has been found to moderate the effects of personality and mobile technology preferences, which, in turn, affect users' willingness to use mHealth apps \cite{nunes2019individual}. In our study, we will utilize structural equation modeling to quantitatively measure how various factors collectively contribute to users' willingness to use mHealth apps, and draw a comprehensive picture of the underlying structure.

\subsection{Approval and Consent}

The experiment was approved by the Institutional Review Board (IRB) of National University of Singapore. A participation information sheet was shown at the first page of the experiment. Participants provided consent before starting the experiment.

\subsection{Participants Recruitment}

We recruited a total of 1,669 participants (1,021 male, mean age 39.64 $\pm$ 7.83) from eight countries (US, UK, Germany, India, China, Singapore, New Zealand, and Australia) across four continents. Among them, 436 participants were from the online crowd-sourcing platform Amazon Mechanical Turk (MTurk) \cite{litman2020conducting}, and 1,233 participants were from Toluna Global Panel, another crowd-sourcing platform managed by the Internet survey company Toluna (www.toluna-group.com). We used two platforms for two reasons: first, MTurk has few participants from Oceania and Asia (except India), so we used Toluna to obtain more participants from eight different countries; second, MTurk has few participants over the age of 60, but we aimed to recruit a significant number of seniors as they are a specific target audience of mHealth apps. Toluna enabled us to recruit seniors as well as participants in other age groups. For each country, we attempted to achieve a uniform distribution of participants' ages. Overall, our participants had diverse demographic backgrounds in terms of their residential countries, gender, and age (see Table \ref{tab:demographic} for details).

\begin{table}[htbp] \small
	\centering
	\caption{Participant Demographics. Distribution for gender, age, and education.}
	\begin{tabular}{cp{4.3cm}rrrrrrrr}
		\hlinew{0.8pt}
		&       & \multicolumn{1}{c}{US} & \multicolumn{1}{c}{IN} & \multicolumn{1}{c}{CN} & \multicolumn{1}{c}{SG} & \multicolumn{1}{c}{UK} & \multicolumn{1}{c}{DE} & \multicolumn{1}{c}{AU} & \multicolumn{1}{c}{NZ} \\
		\hline
		\multicolumn{2}{c}{Number of participants} & \multicolumn{1}{c}{204} & \multicolumn{1}{c}{206} & \multicolumn{1}{c}{315} & \multicolumn{1}{c}{159} & \multicolumn{1}{c}{169} & \multicolumn{1}{c}{111} & \multicolumn{1}{c}{343} & \multicolumn{1}{c}{186} \\
		\midrule
		\multirow{2}[2]{*}{\begin{sideways}\textbf{Gen.}\end{sideways}} & Male  & 60.78\% & 73.30\% & 53.97\% & 62.26\% & 57.40\% & 95.41\% & 50.73\% & 54.84\% \\
		& Female & 39.22\% & 26.70\% & 46.03\% & 37.74\% & 42.60\% & 4.59\% & 49.27\% & 45.16\% \\
		\hline
		\multirow{6}[2]{*}{\begin{sideways}\textbf{Age}\end{sideways}} & Below 19 & 0.49\% & 1.46\% & 0.32\% & 3.77\% & 0.59\% & 0.00\% & 1.75\% & 1.61\% \\
		& 19-24 & 7.84\% & 3.88\% & 4.76\% & 8.81\% & 8.28\% & 32.43\% & 9.62\% & 11.29\% \\
		& 25-34 & 45.10\% & 69.42\% & 30.16\% & 29.56\% & 26.63\% & 37.84\% & 24.78\% & 20.97\% \\
		& 35-44 & 25.00\% & 20.87\% & 24.13\% & 32.08\% & 17.16\% & 5.41\% & 25.36\% & 21.51\% \\
		& 45-59 & 16.67\% & 3.40\% & 30.48\% & 20.75\% & 27.81\% & 1.80\% & 16.62\% & 26.34\% \\
		& Above 60 & 4.90\% & 0.97\% & 10.16\% & 5.03\% & 19.53\% & 29.73\% & 21.87\% & 18.28\% \\
		\hline
		\multirow{6}[2]{*}{\begin{sideways}\textbf{Ethnicity}\end{sideways}} & American Indian & 4.90\% & 3.22\% & 0.00\% & 1.26\% & 0.00\% & 0.00\% & 2.33\% & 1.61\% \\
		& Asian (Indian)/Pacific islander & 8.82\% & 91.49\% & 95.87\% & 88.05\% & 11.24\% & 38.53\% & 16.62\% & 22.04\% \\
		& Black & 4.41\% & 0.69\% & 0.00\% & 1.26\% & 4.14\% & 1.83\% & 1.17\% & 2.69\% \\
		& Hispanic/Latino & 3.43\% & 0.46\% & 0.00\% & 0.63\% & 0.00\% & 0.00\% & 1.75\% & 1.08\% \\
		& White & 75.00\% & 1.84\% & 0.63\% & 4.40\% & 84.02\% & 58.72\% & 74.34\% & 63.44\% \\
		& Other & 3.43\% & 2.30\% & 3.49\% & 4.40\% & 0.59\% & 0.92\% & 3.79\% & 9.14\% \\
		\hline
		\multirow{5}[2]{*}{\begin{sideways}\textbf{Education}\end{sideways}} & Below high school & 2.45\% & 0.97\% & 5.08\% & 3.14\% & 2.37\% & 0.00\% & 4.08\% & 3.23\% \\
		& Vocational training & 9.31\% & 1.94\% & 9.21\% & 8.18\% & 29.59\% & 1.83\% & 12.83\% & 20.43\% \\
		& High school graduate & 5.88\% & 2.43\% & 7.94\% & 18.24\% & 19.53\% & 0.92\% & 28.57\% & 23.12\% \\
		& Bachelor's degree & 59.31\% & 82.04\% & 66.03\% & 59.12\% & 27.81\% & 87.16\% & 34.11\% & 36.56\% \\
		& Graduate degree & 23.04\% & 12.62\% & 11.75\% & 11.32\% & 20.71\% & 10.09\% & 20.41\% & 16.67\% \\
		\hlinew{0.8pt}
	\end{tabular}%
	\label{tab:demographic}%
\end{table}%


\subsection{Vignette Design} \label{sec:vignette}

Our human experiment is based on the vignette design, which combines the advantages of survey research and multidimensional experimental design. In vignette experiments, short, systematically varied descriptions of situations or subjects (called vignettes) are used to elicit the beliefs, attitudes, or behaviors of respondents regarding presented scenarios. Participants evaluate hypothetical situations or subjects described in vignettes that vary in the level of characteristics (dimensions) of the described situations or subjects \cite{steiner2016designing, keusch2019willingness}. The advantage of vignette experiments is that they allow researchers to study complex social phenomena in a controlled and systematic way. By using carefully constructed scenarios, researchers can manipulate different variables to examine their effects on people's attitudes and behaviors. This can help isolate and identify specific factors that influence how people think and act in different social situations \cite{harrits2021qualitative}. 

\subsubsection{Vignette dimensions}

The vignettes used in our study are composed of various dimensions of mHealth apps, inspired by our hypotheses H2 and H3 (see Section \ref{sec:hypotheses}). Each dimension is assigned one of multiple levels, which we determined by examining existing mHealth apps and prior research. We selected a final set of seven factors and factor levels, which are reported in Table \ref{tab:vignette}.


\makeatletter
\def\hlinew#1{%
	\noalign{\ifnum0=`}\fi\hrule \@height #1 \futurelet
	\reserved@a\@xhline}

\begin{table}[htbp] \small
	\centering
	\caption{Vignette dimensions, number of levels, and vignette text used in our human study.}
	\begin{tabular}{|p{3.1cm}|p{1.6cm}|p{10.7cm}|}
		\hlinew{0.8pt}
		\multicolumn{1}{|l|}{\textbf{Dimension}} & \multicolumn{1}{l|}{\textbf{No. of levels}} & \textbf{Vignette text} \\
		\hline
		\multirow{4}[8]{*}{App functionality} & \multirow{4}[8]{*}{4 levels} & provides general lifestyle related suggestions for maintaining a healthy life \\
		\cline{3-3}          &       & provides personalized feedback and suggestions about your personal health \\
		\cline{3-3}          &       & monitors your health and provides personalized feedback \\
		\cline{3-3}          &       & allows your family members or friends to monitor your health remotely and receive real-time alert during medical emergencies \\
		\hline
		\multirow{3}[6]{*}{App rewards} & \multirow{3}[6]{*}{3 levels} & discounts/coupons for health check-ups, specialist/doctor consultancy and medical tests \\
		\cline{3-3}          &       & discounts/coupons for health/fitness products such as smart-watches, exercise equipment and gym memberships \\
		\cline{3-3}          &       & to know and interact with people who have similar health conditions \\
	\hline
		\multirow{3}[6]{*}{Data collection} & \multirow{3}[6]{*}{3 levels} & your daily statistics such as step count, resting heart rate, geolocation, physical movements \\
		\cline{3-3}          &       & your health records such as medication and clinic consultation records, daily diet \\
		\cline{3-3}          &       & your daily usage on mobile phone such as number of messages and phone calls, browser history, app usage, and interaction with other devices \\
	\hline
		\multirow{3}[6]{*}{Data storage} & \multirow{3}[6]{*}{3 levels} & not be stored either on phone or on the cloud \\
\cline{3-3}          &       & be stored on your phone only \\
\cline{3-3}          &       & be stored at the app developer side \\
\hline	
		\multirow{3}[6]{*}{Data sharing} & \multirow{3}[6]{*}{3 levels} & not to be shared \\
\cline{3-3}          &       & be shared with the app developer \\
\cline{3-3}          &       & be shared with the hospitals and clinics \\
\hline	
		\multirow{2}[4]{*}{Data protection} & \multirow{2}[4]{*}{2 levels} & protected by the latest encryption and authentication techniques \\
		\cline{3-3}          &       & No mentioning of the privacy \& security mechanisms \\
	\hline
		\multirow{3}[6]{*}{User's level of control} & \multirow{3}[6]{*}{3 levels} & user can stop data collection and sharing by changing the settings in the app \\
		\cline{3-3}          &       & user can set in the app on which data to be collected or shared \\
		\cline{3-3}          &       & No mentioning of level of control \\
		\hlinew{0.8pt}
	\end{tabular}%
	\label{tab:vignette}%
\end{table}%

\begin{figure*}[ht]
	\centering
	\includegraphics[width=.99\textwidth]{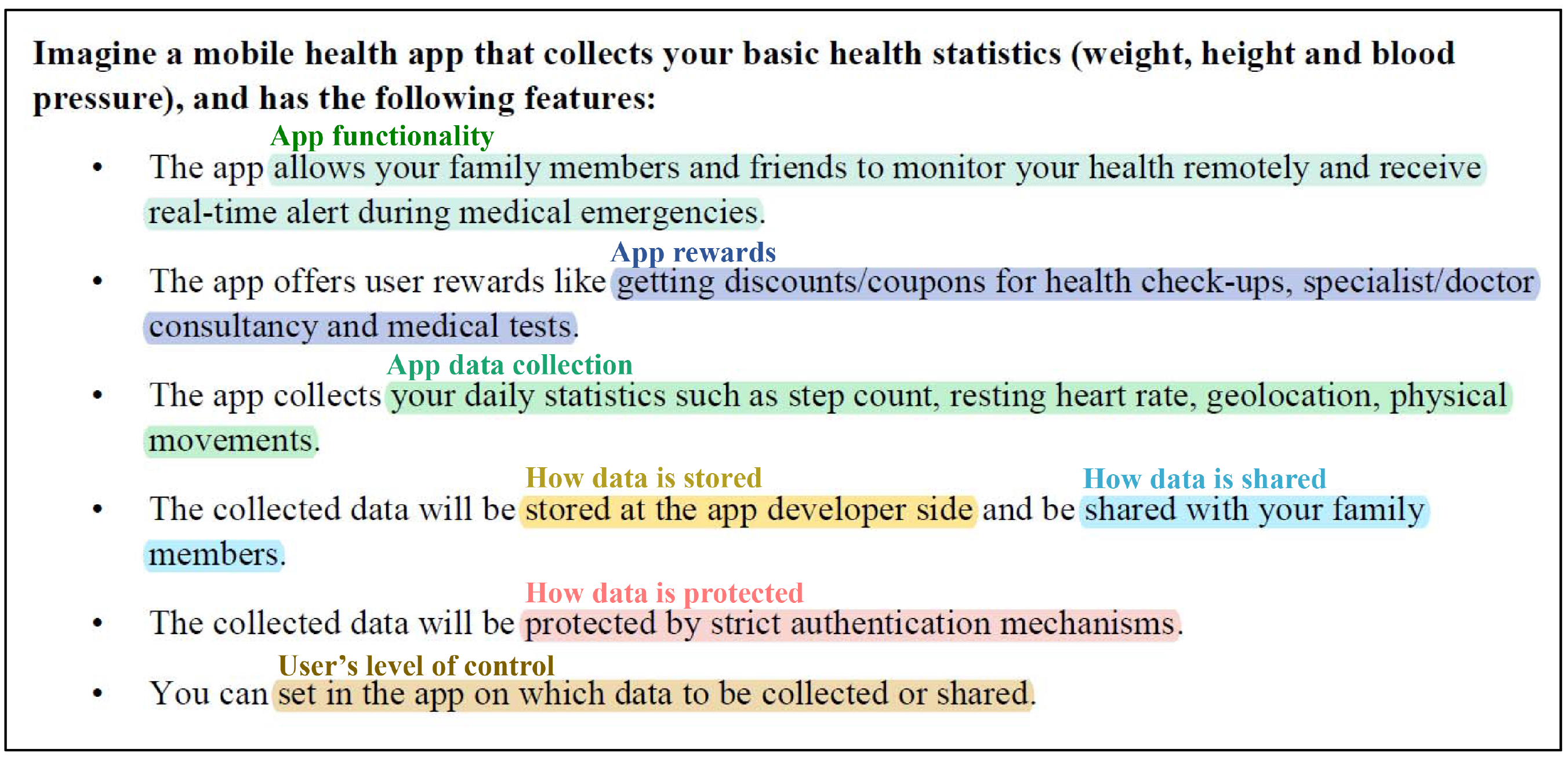}
	\caption{Examples of vignette that combines seven factor levels into a specific scenario of mHealth app.}  
	\label{fig:vignette} 
\end{figure*}

\subsubsection{Vignette composition}

In our study, we used vignettes to describe hypothetical mHealth apps, as illustrated in Fig. \ref{fig:vignette}. Each vignette consisted of an unchanging text template (the non-highlighted black text) with placeholders for six factors (colored boxes), each assigned one of several levels (the text in the colored boxes) to create a unique scenario. We systematically varied the factor levels across various vignettes to evaluate how different levels influenced participants' perception of the hypothetical mHealth apps. In total, we generated 1,944 vignettes by combining all possible factor levels, resulting in a diverse set of scenarios to explore participants' attitudes and behaviors towards mHealth apps.


\subsection{Experiment Procedure} 

Our human study was performed on two crowd-sourcing platforms - Amazon Mechanical Turk (MTurk) and Toluna. We recruited participants from the US and India through MTurk and participants from other countries through Toluna. The vignettes and surveys were translated to Chinese for participants from China, while participants from other countries responded in English. Each participant was presented with eight vignettes, which described a hypothetical mHealth app. Participants rated the likelihood of using the app on a 7-point scale, where 1 indicated "Very unlikely" and 7 indicated "Very likely". They were also asked to provide a reason for their response, rate the perceived usefulness of the app on a scale of 1 to 7, and indicate their major goal if they decided to use the app. Additionally, participants were asked about what they liked most and least about the app and their demographic information, such as gender, age, ethnicity, educational background, and country of residence. Questions were also included about participant information, such as digital literacy and online behavior. All measures used in the study are available in Appendix B.

\subsection{Analytical Approach}

We conducted statistical analyses using several methods to examine the impact of each independent variable on the dependent variables. Firstly, we conducted ANOVA (analysis of variance) to check for any overall effects. If ANOVA results were significant, we performed post-hoc Tukey tests to identify specific manipulation effects. For analyses involving multiple factors, we used both logistic regression and linear regression. To control for multiple comparisons, we applied the Bonferroni correction. For more information about these inferential statistics, please refer to Bailey's (2008) guide \cite{bailey2008design}.

We used structural equation modeling to gain a comprehensive understanding of how multiple factors collectively influence users' behavioral intention. In particular, we first conducted an exploratory factor analysis (EFA) to identify the number of latent variables (factors) present in the human data and extract a concise set of attributes underlying these variables. Next, we conducted a confirmatory factor analysis (CFA) to assess the reliability and validity of the measures and test the relations between latent and observed variables. Factor analysis (FA) and principal component analysis (PCA) are variable reduction techniques that extract a reduced set of variables from a larger set of variables, but in PCA, the components are orthogonal linear combinations that maximize the total variance, while in FA, the factors are linear combinations that maximize the shared portion of the variance underlying "latent constructs." Attributes with poor loadings or fits were removed. We developed the final model through path analysis, which calculated standardized regression weights ($\gamma$) and correlations ($\phi$) among latent variables. We conducted these analyses using statistical toolboxes in Matlab R2016b, IBM Amos 21, and IBM SPSS 20.

\subsection{Data Availability}

Readers can view the vignettes and try our human study online at \url{https://ncript.comp.nus.edu.sg/site/experiment2/start?taskid=4391}. 

\section{Results}

In this section, we present the results of our human study. Specifically, we explored how different factors impact two important ratings: i) participants' willingness to use the app and ii) their perception of its usefulness. Our investigation sought to shed light on the underlying structure that shapes users' behavioral intention to use mHealth apps.

\subsection{Influence of App Related Properties} \label{sec:app}

We conducted ANOVAs for the seven factors listed in Table \ref{tab:vignette} to examine whether each factor influences participants' willingness and perceived app usefulness. Results indicated that three factors (rewards for using the app, the types of data collected by the app, and how data is shared by the app) significantly impacted participants' willingness, with $F$s $\geq 5.13$ and $p$s $\leq .006$\footnote{ANOVA results are presented as "$F$(df${condition}$, df${error}$) = $F$ value, $p$ = $p$ value". If a $p$ value is less than the conventional significance level threshold of $.05$ or $.01$ (with Bonferroni correction), we reject the null hypothesis of no difference among the means.}. Financial rewards (e.g., discount/coupons for health service and products) had greater effect on promoting user willingness than knowing and interacting with people with similar health conditions, with $F(2,14518) = 5.13$ and $p = .006$. The factor of how data is stored had a marginal impact on participants' willingness, with $F(2,14518) = 5.01$ and $p = .007$. App functionality, how the app data is protected, and users' level of control did not have significant impact. Only one factor, the types of data collected by the app, had a significant impact on perceived app usefulness, with $F(2,14518) = 26.13$ and $p \leq .0001$. Detailed post-hoc Tukey test results for the impact of separate factor levels on participants' willingness are provided in Table \ref{tab:appfactor}.

\definecolor{LightCyan}{rgb}{0.54, 0.81, 0.94}
\definecolor{LightOrange}{rgb}{0.94, 0.82, 0.74}

\begin{table}[htbp] \small
	\centering
	\caption{The participants' willingness to use the mHealth app varied depending on different factor levels. For each type of factor, the willingness under the level with a bold font was statistically higher ($p$s $\leq .01$).}
	\begin{tabular}{|p{2.75cm}|p{10.6cm}|p{1.9cm}|}
	\hlinew{0.8pt}
		\multicolumn{1}{|l|}{\textbf{Type of factor }} & \textbf{Factor level} & \multicolumn{1}{l|}{\textbf{Willingness}} \\
	\hlinew{0.8pt}
		\multirow{3}[6]{*}{App rewards} &  \textbf{discounts/coupons for health check-ups, specialist/doctor consultancy and medical tests} &  \textbf{0.69} \\
	\cline{2-3} 		
         &  \textbf{discounts/coupons for health/fitness products such as smart-watches, exercise equipment and gym memberships} &  \textbf{0.69} \\
	\cline{2-3}         

		 & to know and interact with people who have similar health conditions & \ 0.68 \\
	\hlinew{0.8pt}
		\multirow{3}[6]{*}{Data collection} & \textbf{your daily statistics such as step count, resting heart rate, geolocation, physical movements} & \textbf{0.70} \\
	\cline{2-3}          & \textbf{your health records such as medication and clinic consultation records, daily diet} & \textbf{0.70} \\
	\cline{2-3}          &  your daily usage on mobile phone such as number of messages and phone calls, browser history, app usage, and interaction with other devices & \ 0.66 \\
	\hlinew{0.8pt}
		\multirow{3}[6]{*}{Data sharing} & \textbf{not to be shared} & \textbf{0.70} \\
	\cline{2-3}         &  be shared with the app developer &  \ 0.68 \\
	\cline{2-3}          & be shared with the hospitals and clinics &  \ 0.68 \\
	\hlinew{0.8pt}
	\end{tabular}%
	\label{tab:appfactor}%
\end{table}%

The ANOVA revealed interaction effects of country of residence with app rewards and data collection on users' willingness, with $F$s $\geq 2.16$ and $p$s $\leq .007$. Follow-up analyses showed that the effect of app rewards was primarily driven by German participants ($F(2,1004) = 5.23$, $p = .006$) and New Zealand participants ($F(2,1467) = 5.17$, $p = .006$), while the effect of data collection was significant among most countries except India, China, and the United Kingdom, with $F$s $\geq 6.78$ and $p$s $\leq .001$. No interaction effect of data collection and country of residence was found for perceived app usefulness. 

\begin{figure*}[ht]
	\centering
	\includegraphics[width=.9\textwidth]{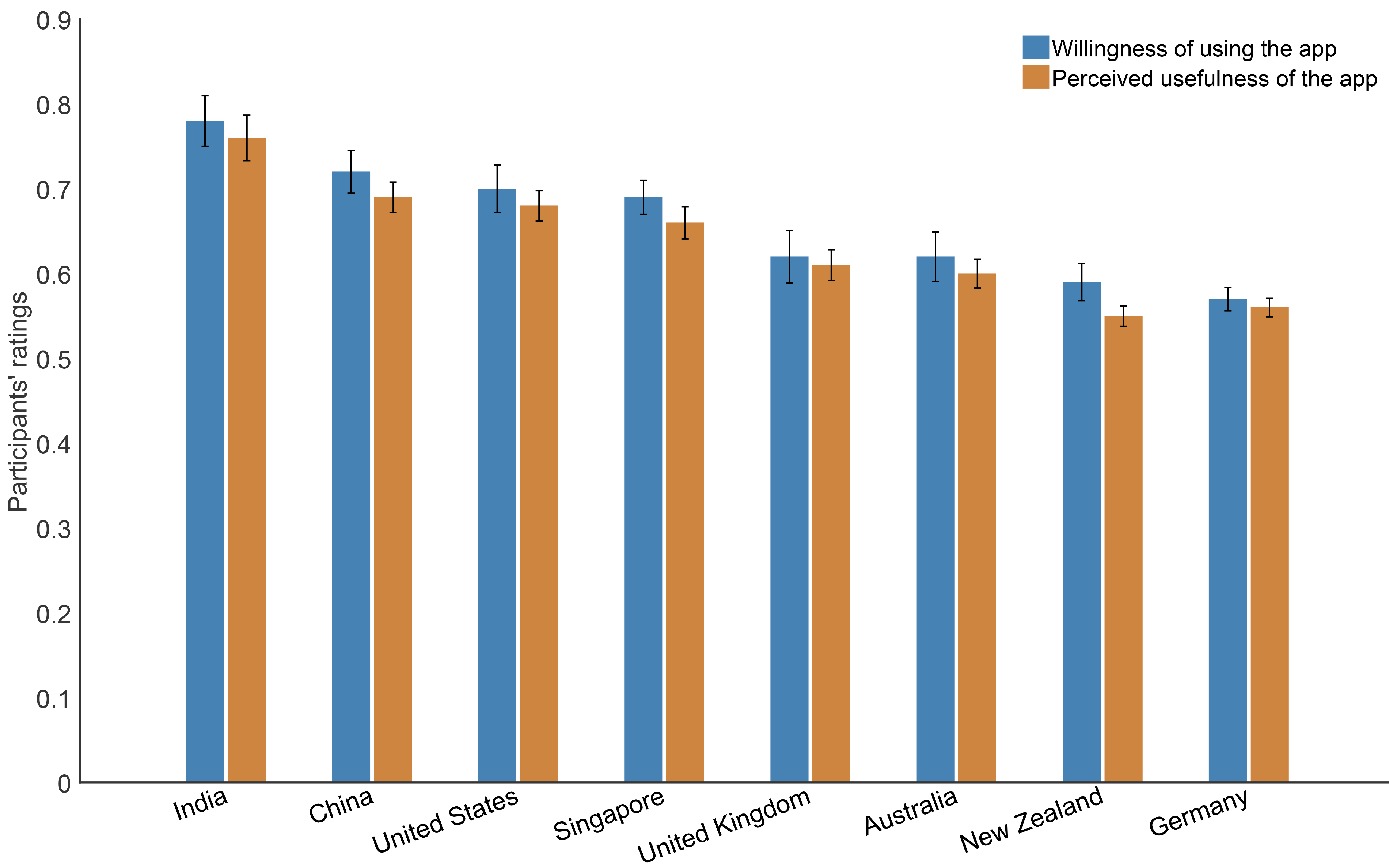}
	\caption{Participant ratings ($\pm$ $SE$) on willingness to use the mHealth app and perceived app usefulness, grouped by country of residences.}  
	\label{fig:comparecountry} 
\end{figure*}

\begin{figure*}[ht]
	\centering
	\includegraphics[width=.99\textwidth]{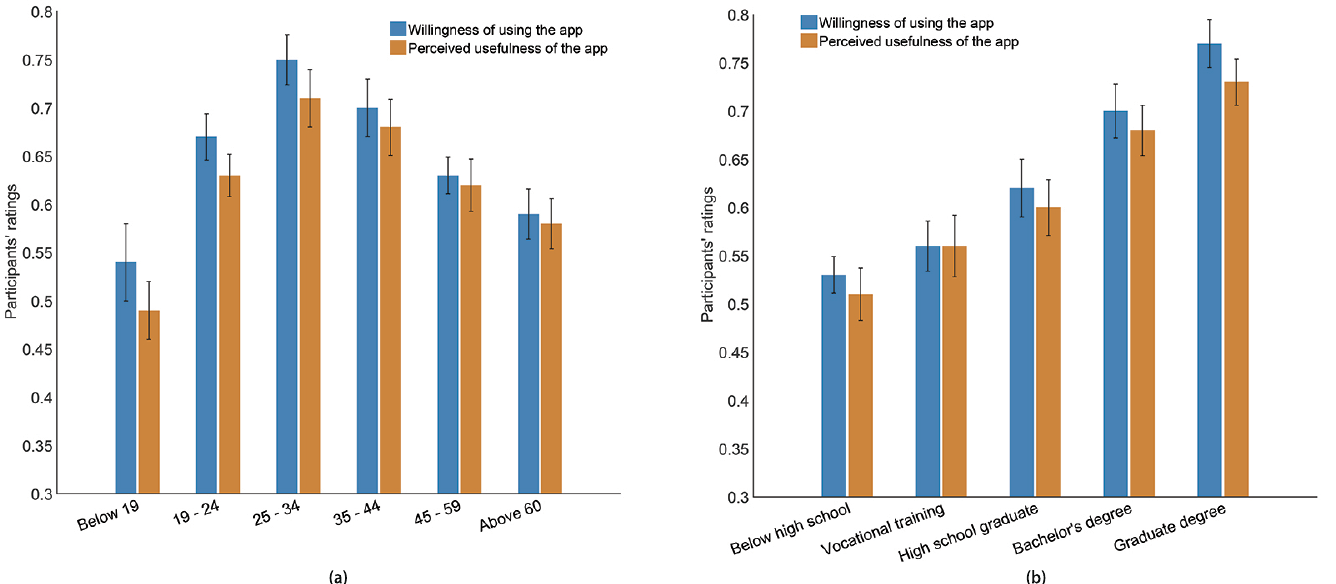}
	\caption{Participant ratings ($\pm$ $SE$) on willingness to use the mHealth app and perceived app usefulness, grouped by ages (a) and educational background (b).}   
	\label{fig:ageedu} 
\end{figure*}

\subsection{Influence of Users' Demographic Background} \label{sec:demo}

In this section, we will explore the impact of participants' demographic backgrounds on their willingness to use the app and their perceived usefulness of it. We will be focusing on five key demographic factors across all app scenarios: country of residence, gender, age, ethnicity, and educational background.

The ANOVA indicated a significant effect of country of residence on both user willingness to use mHealth apps ($F(7,14513) = 116.08$, $p \leq .0001$) and perceived app usefulness ($F(7,14513) = 147.29$, $p \leq .0001$). Indian participants had the highest willingness to use the apps, followed by Chinese and American participants. Participants from Germany and New Zealand showed the lowest intention to use mHealth apps ($p \leq .05$). Participants' perceived usefulness of the app followed a similar pattern (see Fig. \ref{fig:comparecountry}).

The age of the participants had a significant impact on both their willingness to use the mHealth apps ($F(7,14515) = 387.64$, $p$ $\leq .0001$) and their perceived usefulness of the app ($F(7,14515) = 132.63$, $p$ $\leq .0001$), as illustrated in Figure \ref{fig:ageedu}(a). The distribution of age over user willingness formed a bell curve, with participants in the age group of 25-34 exhibiting the highest willingness to use the app and perceiving it to be the most useful. The second-highest willingness and perceived usefulness were demonstrated by the age group of 35-44. Conversely, participants aged below 19 and above 60 showed the lowest willingness to use the app and perceived its usefulness to be the least ($p$s $\leq .05$).

Notably, participants in the age group of 45-59 reported a similar perceived usefulness of the app as those in the age group of 19-24, but the former group demonstrated a lower willingness to use the app compared to the latter. Additionally, participants aged 35-44 rated the app's usefulness higher than those in the age group of 19-24, yet both groups exhibited statistically similar intention to use the mHealth app. These findings suggest that other factors, such as technical skills and related experience, may influence users' behavioral intention. Therefore, a more comprehensive analysis is necessary, as discussed later in Section \ref{sec:joint}.

Participants' educational background and ethnicity also had significant impacts on their willingness to use the app and perceived app usefulness ($F$s $\geq 87.49$, $p$ $\leq .0001$ for ethnicity, and $\geq 182.49$, $p$ $\leq .0001$ for education). Those with higher levels of education were more willing to use the app and perceived it to be more useful (see Fig. \ref{fig:ageedu}(b)). Asians (including Asian Indians) and American Indians showed greater willingness to use the app than Black people, while White and Hispanic participants were the least willing. Gender did not significantly influence users' willingness to use the app ($t(14519) = 1.80$, $p$ $= .071$), but male participants perceived the app as more useful than female participants (Male: $.65$, Female: $.63$, $t(14519) = 4.90$, $p$ $< .0001$). "



\subsection{Influence of Other User Background}

We conducted exploratory analyses on individual difference measures to determine which factors predicted users' willingness to use the mHealth apps. To do this, we calculated the average willingness of each participant to use the eight different versions of the mHealth apps described in the vignettes. We then performed a multiple linear regression analysis using the participants' willingness as the dependent variable and individual difference measures as explanatory variables.

Table \ref{tab:regression} shows that variables related to privacy and security concerns, online behaviors, prior experience of related apps and the Covid-19 pandemic, and technology skills robustly impacted users' willingness to use mHealth apps. The strongest predictor was users' smartphone skills, followed by users' positive attitude towards wearable devices, and users' frequency of sharing health information online. Surprisingly, whether a user had a chronic disease or whether they had used mHealth apps before did not have a significant impact.

\begin{table}[htbp] \small
	\centering
\begin{threeparttable}[b]
	\caption{Regression $\beta$ coefficients for individual difference variables predicting users' willingness of using our mHealth apps. Results of variables with significant impact are highlighted in bold. Bonferroni correction was applied with a reduced $\alpha$ of $.005$.}
	\begin{tabular}{p{10.75cm}p{2cm}p{2cm}}
	\hlinew{0.8pt}
		\textbf{Explanatory variables} & \multicolumn{1}{p{5.785em}}{\textbf{$\beta$}} & \multicolumn{1}{p{2cm}}{\textbf{$p$}} \\
		\hline
		\multicolumn{3}{p{10.75cm}}{Concern about personal privacy and security} \\
		\hline
		Concern about privacy breach on mobile apps & \textbf{-.076} & \textbf{.000} \\

		Knowledge of online info misuse  & \textbf{-.076} & \textbf{.000} \\

		Overall concern about personal privacy & \textbf{-.044} & \textbf{.005} \\

		Worry about companies having access to my profile & \textbf{-.072} & \textbf{.000} \\

		Worry of mobile info leakage & \ -.030 & \ .037 \\

		Worry of info leakage on mHealth apps & \textbf{-.107} & \textbf{.000} \\
	
		Concern about sharing health info online & \textbf{-.107} & \textbf{.000} \\

		Privacy issues and my mobile data activities are not a concern & \ \ .016  & \ .222 \\
	
		Making transactions on my mobile phone is not a concern & \textbf{\ .042} & \textbf{.000} \\


		IUIPC Privacy Concern measurement\tnote{\**} & \textbf{-.084} & \textbf{.000} \\
		\hline
		\multicolumn{3}{p{10.75cm}}{Online behaviors} \\
		\hline
		Daily time of using cell phone & \ -.024 & \ .017 \\

		Freqency of sharing personal info online & \ \ .030  & \ .140 \\

		Frequency of sharing health info online & \textbf{\ .133} & \textbf{.000} \\
		\hline
		Prior experience \& opinions of related apps &       &  \\
		\hline
		Have used mHealth apps before & \ \ .023  & \ .127 \\

		Have used wearable devices before & \textbf{-.037} & \textbf{.003} \\
		Will recommend mHealth apps & \textbf{\ .044} & \textbf{.001} \\
		
		Will recommend wearable Devices & \textbf{\ .139} & \textbf{.000} \\
		
		Acceptance level of tracing app & \textbf{\ .045} & \textbf{.000} \\
		
		Acceptance level of sharing data with public authorities for Covid-19 measures & \textbf{\ .041} & \textbf{.000} \\
		
		\hline
		Related experience about Covid-19 and chronic diseases &       &  \\
		\hline
		Concern about Covid-19 & \textbf{-.055} & \textbf{.000} \\

		Have chronic disease & \ -.003 & \ .573 \\

		Have been quarantined for Covid-19 before & \textbf{\ .038} & \textbf{.000} \\

		Have infected with Covid-19 before & \ -.021 & \ .016 \\

		Have friends infected with Covid-19 before & \textbf{\ .054} & \textbf{.000} \\
    	\hline
		\multicolumn{3}{p{10.75cm}}{Technology skills} \\
		\hline
		Smartphone skills & \textbf{\ .244} & \textbf{.000} \\

		Tech savvy  & \textbf{\ .083} & \textbf{.000} \\

		Technophobia & \textbf{\ .103} & \textbf{.000} \\
	\hlinew{0.8pt}
	\end{tabular}%
\begin{tablenotes}
	\footnotesize
	\item \** Internet Users' Information Privacy Concerns (IUIPC) privacy concern measurement scale \cite{malhotra2004internet}. We report the mean results here.
\end{tablenotes}
\label{tab:regression}
\end{threeparttable}
\end{table}%

\begin{figure*}[ht]
	\centering
	\includegraphics[width=.99\textwidth]{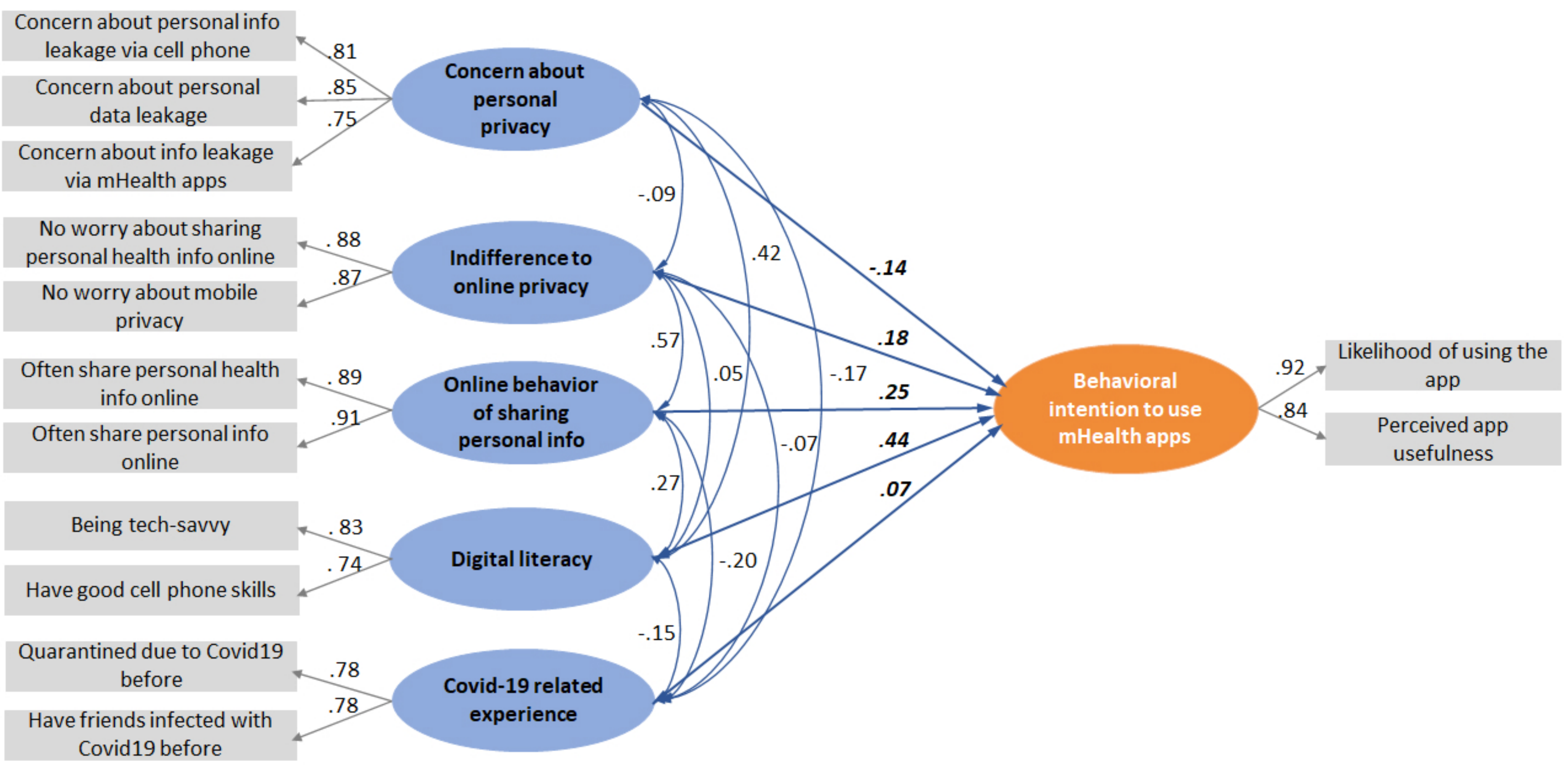}
	\caption{A Structural equation model (SEM) analysis of multiple latent factors (in blue eclipses) contributing to users' behavioral intention to use mHealth apps (in orange eclipse). Individual difference measures describing each latent factor are in grey rectangles. Standardized regression weights ($\gamma$) are in bold italic font. $CFI = .983$, $RMSEA = .045$.}  
	\label{fig:SEM} 
\end{figure*}

\subsection{Joint Influence of Multiple Impacting Factors} \label{sec:joint}

In this section, we utilized structural equation modeling (SEM) to gain a comprehensive understanding of how users' behavioral intention towards our mHealth apps is influenced by various factors. While our previous analysis using multiple linear regression provided insight into the individual impact of variables, SEM allows us to explore the intercorrelations among these variables and how they jointly shape user willingness.

To begin, we conducted exploratory factor analysis (EFA) followed by confirmatory factor analysis (CFA) \cite{kline2015principles} to measure the relationships between observed variables and latent factors (higher-level perceptions and reactions). We hypothesized that the reasons for user willingness are multidimensional and inter-correlated, so we used maximum likelihood with oblique transformation in our EFA \cite{kim1978introduction}. Finally, we employed path analysis to inform the model structure. 

The resulting model, shown in Fig. \ref{fig:SEM}, is divided into three layers. The individual difference measures obtained from our questionnaire are at the bottom layer (in grey rectangles) and load onto different latent variables represented by blue eclipses in the mid-layer. The top layer contains the latent variable ``behavioral intention to use mHealth apps'', represented by an orange eclipse. We created the final model through path analysis, predicting the top layer latent construct from the lower-level perception latent constructs.

We evaluated the model's fitness using two metrics: the Comparative Fit Index (CFI) and Root Mean Square Error of Approximation (RMSEA). Our model had acceptable fit, with $CFI = .983$ and $RMSEA = . 045$. CFI compares the target model's chi-square to an independent model, while RMSEA estimates the error of approximation per model degree of freedom and considers the sample size. Higher CFI and lower RMSEA values indicate better model fit.

In our SEM analysis, as depicted in Fig. \ref{fig:SEM}, five latent factors (represented by blue eclipses) were found to jointly contribute to users' behavioral intention to use mHealth apps. Among these, ``digital literacy'' had the strongest weight on users' behavioral intention ($\gamma = .44$), followed by ``Online behavior of sharing personal information'' ($\gamma = .25$), ``Indifference to personal privacy'' ($\gamma = .18$), and ``Concern about personal privacy'' ($\gamma = -.14$). The factor ``Covid-19 related experience'' had the lowest weight on users' behavioral intention ($\gamma = .07$).

The individual difference measures in our study were moderated by users' demographic background, including age, gender, ethnicity, education, and country of residence. As discussed in Section \ref{sec:app}, we found significant interaction effects of country of residence with app rewards and data collection. Gender also interacted with several measures on users' willingness, such as ``having good cell phone skills'' (gender: $F(6,14507) = 18.61$, $p \leq .0001$) and ``concern about personal info leakage via mHealth apps'' (gender: $F(4,14511) = 7.28$, $p \leq .0001$). Specifically, the impact of ``having good cell phone skills'' on users' behavioral intention had a stronger effect on female participants ($\eta_{p} = .21$) than male participants ($\eta_{p} = .11$). In contrast, the influence of ``concern about personal info leakage via mHealth apps'' was carried by male participants ($F(4,8786) = 18.61$, $p \leq .0001$), with no significance found among female participants ($F(4,5725) = 1.23$, $p = .296$).

Furthermore, we observed a significant interaction effect between education and ``being tech-savvy'' on users' willingness ($F(16,14496) = 13.79$, $p \leq .0001$). For example, the impact of ``being tech-savvy'' had a stronger effect size for participants with education below high school ($\eta_{p} = .64$) than for those with a graduate degree ($\eta_{p} = .21$).

\begin{figure*}[ht]
	\centering
	\includegraphics[width=.99\textwidth]{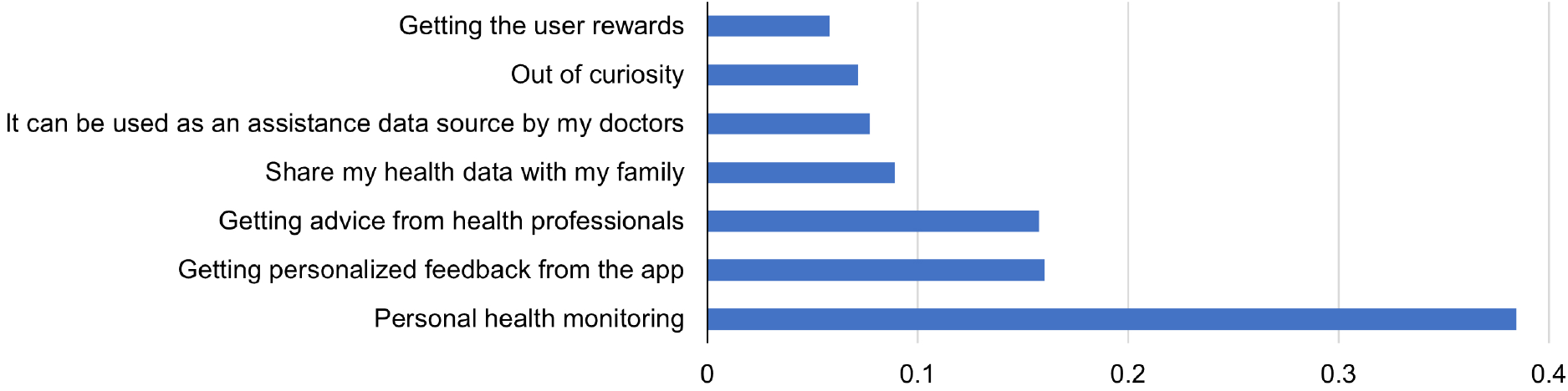}
	\caption{The percentage of options for which participants selected as the major goal of using the mHealth app.}  
	\label{fig:majorgoal} 
\end{figure*} 

\subsection{Users' Purposes and Preferences} \label{sec:pref}

In this section, we aim to investigate participants' usage purposes and preferences for mHealth apps to gain a better understanding of users' behavioral intention. The majority of participants selected ``personal health monitoring'' as their primary goal for using the proposed mHealth apps, followed by ``getting personalized feedback from the app'' and ``receiving advice from health professionals''. Conversely, the least chosen option was ``getting user rewards of using the app'' (see Fig. \ref{fig:majorgoal}). This trend was consistent across all eight countries.

\begin{figure*}[ht]
	\centering
	\includegraphics[width=.99\textwidth]{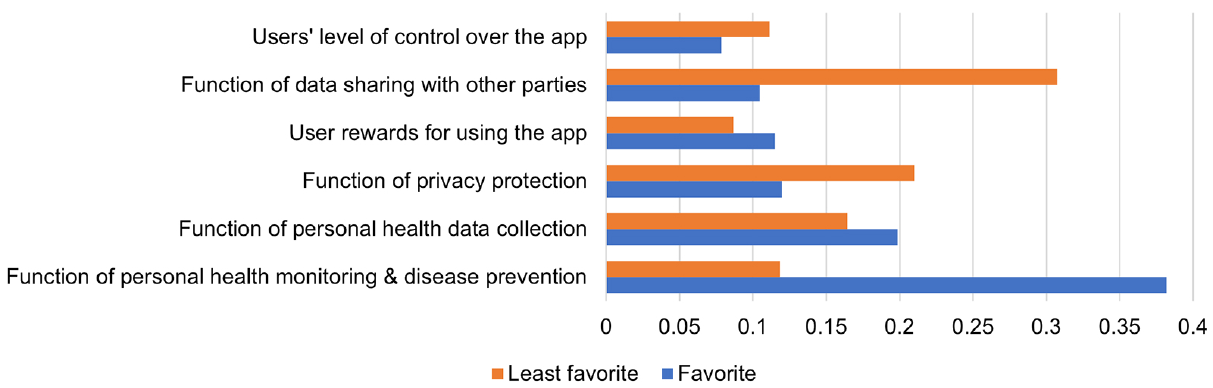}
	\caption{The percentage of options for what participants selected as their favorite and least favorite aspects of the mHealth app.}  
	\label{fig:like} 
\end{figure*}

To gain further insight into participants' opinions, we asked them to identify what they liked most and least about the proposed app separately. As depicted in Fig. \ref{fig:like}, participants appreciated the function of ``personal health monitoring \& disease prevention'' the most and expressed their dislike for data sharing with other parties. Participants showed relative indifference to ``users' level of control over the app'' and ``user rewards for using the app'' in both questions. This trend was consistent across all countries, with the exception of Chinese participants, who indicated that their least favorite feature was privacy protection instead of data sharing with other parties. Participants also provided reasons for their motivation to use or not to use the app, which were consistent with their preferred and least preferred features of the proposed app. For further details, readers can refer to Appendix A.


\section{Discussion}

This section presents a discussion of our findings in relation to previous literature. We also provide a summary of the implications of our results.

\subsection{A General Picture of Users' Behavioral Intention} 


Our findings suggest that the development of users' behavioral intention to use mHealth apps has a complex and hierarchical underlying structure. Multiple factors, such as users' demographic background and technology skills, and the design of the mHealth app, jointly influence users' willingness. 

Based on our SEM, users' digital literacy had the strongest impact, suggesting the critical role of technical competence in promoting mHealth apps. Prior research has suggested the importance of digital technology in digital health \cite{dunn2019technology,grundy2022review}. Our study provides nuanced information that supports prior literature, demonstrating the incomparable importance of digital literacy over other factors in influencing users' behavioral intentions to use mHealth apps. Users' online behavior of sharing personal information online is the second important factor. This is reminiscent of prior research showing the close relation between social network and mobile health \cite{valente2010social,grajales2014social}.

The existing literature has emphasized data privacy and security as a major hindrance to user acceptance of mHealth apps \cite{abelson2017barriers,zhou2019barriers}. Nevertheless, our study reveals that users' privacy concern had only a moderate impact, which was outweighed by users' digital literacy.

Notably, we did not observe a significant correlation between users' privacy concerns and their online behavior of sharing personal information ($p = .999$). This lack of correlation may be attributed to the different underlying factors driving these two latent variables. Users' concern about privacy may arise from a range of reasons, including previous experiences of data breaches or awareness of the potential risks of personal information leakage \cite{huang2022view}. Conversely, users may share their information online due to various reasons, such as convenience, trust in the website or app they are using, or a lack of understanding of the potential consequences of divulging personal information \cite{ahern2007over}. As a result, users may express concerns about privacy but still choose to share their information if they perceive the benefits to outweigh the risks or if they do not fully comprehend the risks involved \cite{anic2019extended}. This observation of a disconnect between users' expressed privacy concerns and their online behavior aligns with the privacy paradox, which refers to the discrepancy between users' intention to protect their privacy and their actual behavior in the online environment \cite{gerber2018explaining}. This explanation also partly accounts for the relatively low weight of users' privacy concerns in their behavioral intention to use mHealth apps.

Ernsting and colleagues \cite{ernsting2017using} discovered that individuals with chronic conditions were more likely to use mHealth apps. However, our study showed that whether participants had a chronic disease had no significant impact on their willingness to use mHealth apps. Similarly, our SEM revealed that users' Covid-19 related experience had a low weight on their behavioral intention. The low correlations observed in our study could be attributed to several factors. Firstly, people's decisions to use mobile health apps are influenced by various factors that may not be directly related to their health status \cite{zaman2022exploring,amagai2022challenges}. Secondly, our SEM indicated that users' digital literacy was a strong influencing factor. However, senior participants who are more likely to have chronic diseases are typically less digitally literate, which hinders them from using mHealth apps \cite{wildenbos2019mobile}. Additionally, the functionality of mHealth apps can be a crucial factor in users' decisions, but the design of the app's features in our vignettes was not tailored to chronic diseases. Previous studies have highlighted the challenges of aging in the context of apps for chronic diseases \cite{wildenbos2019mobile}. Our findings support the need for promoting digital literacy among patients with chronic diseases, as emphasized in prior literature.

\subsection{The Importance of mHealth App Design}

Our analyses showed significant impact for three app related factors: the rewards of using the app, types of data collected by the app, and with whom the data is shared. Our participants cared more about financial rewards (\emph{e.g.}, discount/coupons for health service and products) than knowing and interacting with people with similar health conditions. The effect of financial incentives is consistent with previous studies in \cite{tran2022use}. However, our research only analyzed short-term acceptance. In terms of continued usage of mHealth apps, other factors may came into play, such as performance expectancy and price value \cite{wu2022factors}. 

Our study found that participants were more willing to allow mHealth apps to collect their health records, such as medication and clinic consultation records, than their daily usage data on mobile phones, such as the number of messages and phone calls, browser histories, and app usages. This preference may be due to the clear linkage between health records and mHealth apps, while the connection between the app and daily usage on mobile phones is unclear. However, previous studies have shown that daily phone usage can be informative of a person's mental health \cite{jalali2013building,jameel2022mhealth,marciano2022digital}. We anticipate challenges in raising user awareness and acceptance of mHealth apps that are related to mental health.

Furthermore, our study found that participants' intention to use mHealth apps was negatively influenced by sharing app data with third parties such as app developers and hospitals. Previous research has identified multiple factors that influence users' willingness to share self-collected health data, including the source and type of data, user benefits of data sharing (such as personalized feedback), and privacy and security concerns \cite{woldaregay2020user,kumar2020mobile,grande2022consumer}. To encourage the sharing of mobile health data for the benefit of individuals and the wider community, greater efforts are needed to address these concerns and promote the benefits of data sharing. 

Our analysis did not find a significant influence of app functionality, but this does not necessarily imply that functionalities lack significance. It is possible that the four levels on our functionality dimension are not sufficient to represent the range of functionality offered by mHealth apps in general. Further research should consider more detailed and specialized functionality features that cater to specific groups, such as pregnancy apps designed for women. Such specialized functionality may have a greater impact on users' behavioral intentions to use mHealth apps than general functionality measures. In addition, past studies have indicated that there is a significant correlation between higher levels of perceived ease of use and the intention to use a product or service \cite{venkatesh1999creation}. Our forthcoming research will also explore the significance of user-friendly attributes as a crucial factor in this regard.

\subsection{The Role of Demographic Background} 

Our analysis revealed that users' willingness to use mHealth apps was significantly influenced by their demographic features, such as country of residence, age, and education.

Participants from India, China, and the United States exhibited the highest intention to use the apps, with Singapore following closely behind. In contrast, participants from Europe (United Kingdom and Germany) and Oceania (Australia and New Zealand) showed relatively lower intention compared to those from Asia and the United States. Moreover, the intention of participants from Germany and New Zealand was influenced by app rewards, while those from other countries were not. The type of data collected by the app also significantly influenced the intention of participants from most countries, except India, China, and the United Kingdom.

Several factors could account for these differences. Cultural attitudes towards healthcare and data sharing, for instance, may vary across regions, with some cultures emphasizing self-care and being more open to data collection and sharing. For instance, research by Utz and colleagues \cite{utz2021apps} found that Chinese participants preferred the collection of personalized data, while German and US participants favored anonymity. In some regions, such as India and China, limited access to healthcare services in certain regions could make mHealth apps a more attractive option for individuals seeking to manage their health \cite{okolo2021cannot,yang2021systematic}. Differences in technology adoption and digital literacy could also play a role, with some populations being more comfortable with using mobile devices and apps in their daily lives \cite{martin2006digital}. Finally, varying regulatory frameworks and policies related to mobile health apps could also affect acceptance levels \cite{ashrafi2019comparative,zeller2019right}.

Different from previous research \cite{nunes2019individual,virella2019mobile}, our study identified a bell curve pattern that linked user age and their willingness to use mHealth apps. We found that participants aged between 25 and 34 expressed the strongest intention to use these apps, followed by the age group of 35-44. Participants over the age of 45 or under the age of 25 demonstrated a lower behavioral intention. In particular, participants above the age of 60 reported the lowest willingness to use the apps. Education enhances users' willingness to use mHealth apps. Our SEM analyses suggest that one possible explanation for this trend is that individuals with higher levels of education may possess greater digital literacy.

Our study provides strong evidence that several demographic factors, such as country of residence, age, gender, and education, play a significant role in moderating mHealth adoption. While previous literature has highlighted age and gender as commonly recognized moderators \cite{hoque2016empirical,nunes2019acceptance}, our research offers more nuanced insights by revealing additional moderating factors, including education and country of residence. 

\subsection{Implications of Findings}

Our research may have relevance across various disciplines. First and foremost, our findings can aid mHealth developers in creating more user-welcomed apps. We recommend that designers consider the factors that have a significant impact on users' willingness to use the app, such as the type of data collected by the app and with whom it is shared. Users would like to be informed about the rewards for using the app, but they are less concerned about the app's data protection and storage or the level of control they have over it.

The most significant finding of our SEM analyses is the critical role of users' digital literacy. As the reliance on digital tools increases, it has the potential to exacerbate existing health disparities by widening the gap between those who possess digital skills and access to digital tools and those who do not. In conjunction with the importance of data collection and sharing highlighted in the previous paragraph, our research provides support for policymakers to regulate the data management of mHealth apps and promote digital literacy among users.

Our study provides valuable insights for both healthcare professionals and marketers interested in promoting and implementing mHealth interventions. By demonstrating the significant moderating effects of demographics, our research highlights the importance of tailoring digital health solutions to different populations by considering a wider range of demographic factors. These findings have important implications for improving the effectiveness and accessibility of mHealth interventions, ultimately helping to improve healthcare outcomes for diverse populations.

Finally, our study also presents an invitation for researchers to delve deeper into the underlying structure that shapes users' behavioral intention to use mHealth apps. Through this exploration, we hope to gain a better understanding of how different factors, such as users' demographics and online behavior, impact user willingness to utilize mHealth apps. This deeper understanding will allow for more comprehensive explanations of user behavior and pave the way for improved mHealth app development and adoption.

\section{Limitation and Future Work}

Although our research findings are insightful, there are some limitations to our study. Prior research has emphasized the significance of app cost and perceived ease of use \cite{krebs2015health,liu2020use}, particularly in developing countries and among individuals with low digital literacy \cite{kruse2019barriers}. Unfortunately, we did not measure these two factors in our study, mainly due to two reasons. Firstly, previous research has already produced consistent and robust findings regarding these factors. Secondly, they are primarily dependent on the app developers' design and marketing strategy. Additionally, the app developer's brand image and trustworthiness can also influence users' behavioral intention, which is an essential factor \cite{mathews2019digital}. However, to simplify our experiment's design, we chose to control for it.

As with many studies, our data were collected based on experimentally controlled stimuli and participant self-reports, rather than real-life decisions to use mHealth apps. Although our careful manipulation of variables enhanced the internal validity of our findings, it may limit the generalizability of the results to natural behavior. It is important to acknowledge this limitation and consider the potential implications for real-life scenarios. Nonetheless, our findings provide valuable insights into the underlying mechanisms that influence individuals' decisions to use mHealth apps.

Lastly, it is important to acknowledge that our study's participants were recruited solely from two online platforms (MTurk and Toluna), which may have limited the diversity of our user population. For instance, our participants were all familiar with online crowd-sourcing tasks, indicating that they may possess higher levels of digital literacy than the general population. We tried to include countries that are representative of each continent. However, due to budget constraints and limited participant availability on MTurk and Toluna, our coverage could not be more comprehensive. Moreover, we conducted surveys in English for all countries except China, resulting in a predominantly English-speaking participant pool. Previous studies have suggested that language preferences can significantly impact user perceptions of mHealth apps \cite{liu2020use}, which means that our findings may not be readily generalizable to non-English speaking users.

Moving forward, we intend to expand our participant pool by recruiting individuals from local communities and non-English speaking countries. By doing so, we aim to enhance the generalizability of our study and obtain a more diverse sample of the population. Additionally, we plan to explore the use of a real app setting in future studies. Participants will be asked to install and utilize the app for a set period, and provide feedback on their perceptions and experiences at various stages throughout the trial period. By utilizing a real app setting, we hope to better capture the nuances of user behavior and generate more comprehensive insights into mHealth app usage.

\section{Conclusion}

Mobile health (mHealth) apps have been recognized as a promising solution to improve healthcare outcomes, increase access to care, and reduce healthcare costs \cite{sim2019mobile}. However, their widespread adoption is hindered by barriers such as data privacy and security concerns \cite{zhou2019barriers}. It is crucial to address these barriers to promote the use of mHealth apps. This study offers a comprehensive understanding of the factors that influence users' intention to use mHealth apps and measures their contribution to user willingness. It highlights the critical role of user digital literacy and emphasizes the importance of tailoring solutions to different populations based on a wider range of demographic factors.

The findings of this study have implications for various stakeholders, including app designers, healthcare practitioners, and policymakers. Future research could combine the results of this study with media and sociology research to enhance user willingness to use mHealth apps. In addition, promoting digital literacy and regulating data collection and sharing are crucial for increasing user trust in mHealth apps. Overall, this study provides valuable insights for improving the design and implementation of mHealth apps and promoting their use.

\begin{acks}
This research is supported by the National Research Foundation, Singapore under its Strategic Capability Research Centres Funding Initiative. Any opinions, findings and conclusions or recommendations expressed in this material are those of the author(s) and do not reflect the views of National Research Foundation, Singapore.
\end{acks}

\bibliographystyle{ACM-Reference-Format}
\bibliography{sample-base}

\section*{Appendix}

\appendix



\section{Participants' Self-reported Reasons}

\begin{figure*}[ht]
	\centering
	\includegraphics[width=.85\textwidth]{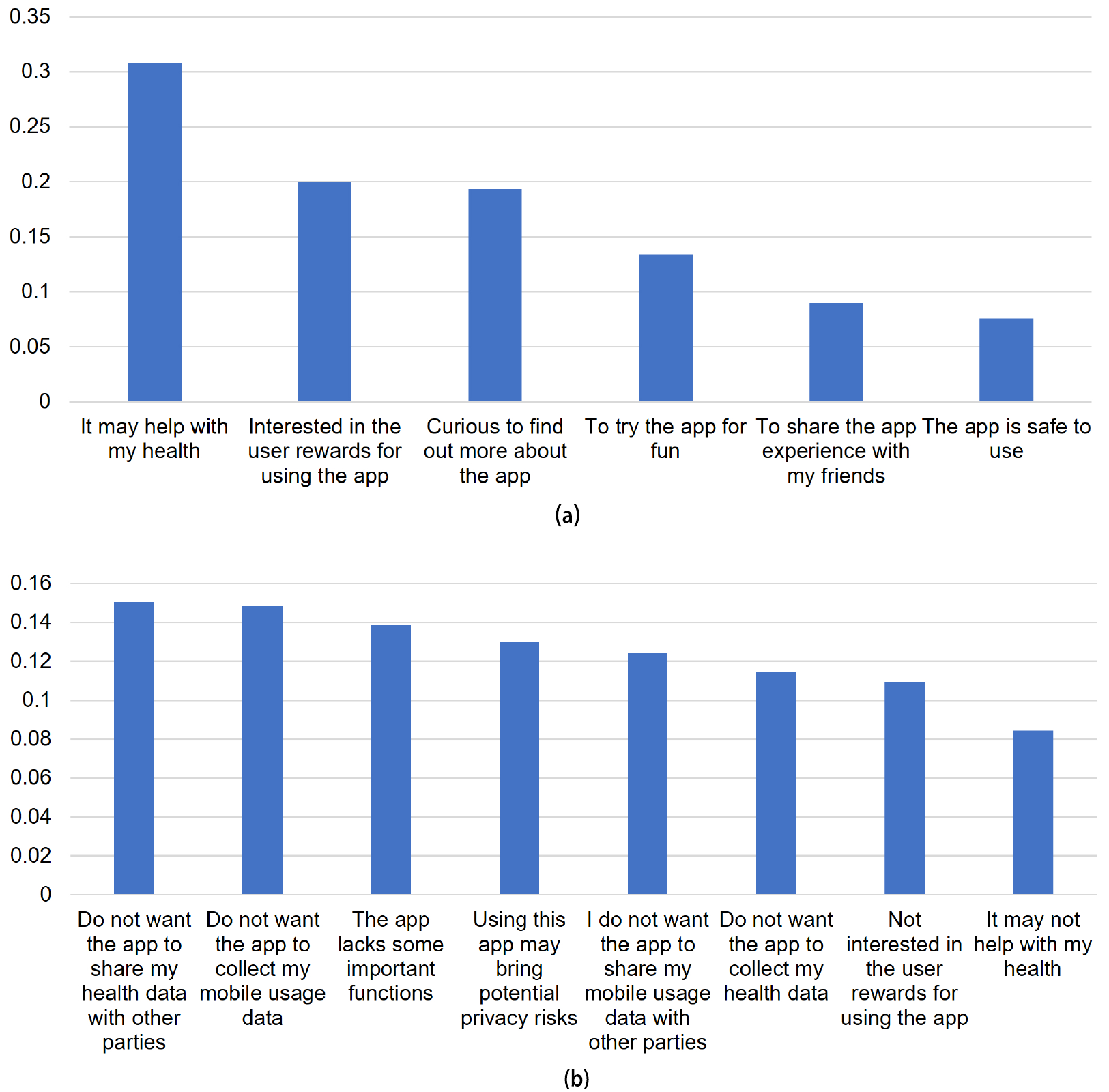}
	\caption{(a) Percentage of participants who selected each reason as a motivating factor for using the mHealth app. (b) Percentage of options chosen by participants for each reason that prevented them from using the mHealth app.}  
	\label{fig:reason} 
\end{figure*}

After each vignette, participants were asked to indicate their reasons for using or not using the proposed mHealth app. Results showed that participants were most motivated to use the app for health-related reasons (see Figure \ref{fig:reason}). However, the top reason preventing usage was ``Do not want the app to share my health data with other parties'', followed by ``Do not want the app to collect my mobile usage data''. This is consistent with the finding in Sec. 4.5 that these options were least favored by participants. The third-highest reason preventing usage was ``The app lacks some important functions'', highlighting the importance of app functionality. Notably, app rewards ranked second in the reasons motivating usage but were among the least selected items for preventing users from using the app. This indicates that app rewards had a significant impact on promoting app usage but did not significantly hinder user acceptance.

\section{Questionnaire}

Section I. You will see 8 different mobile health apps. The bullet items may be different among the apps. The difference with the previous app will be highlighted in BLUE. Please read each app introduction carefully and let us know your perception about them by answering the questions. At the end of this section, there will be questions to validate if you have read carefully.\\

\emph{Sample scenario}: Imagine a mobile health app that collects your basic health statistics (weight, height and blood pressure), and has the following features:

\begin{itemize}
	\item The app provides general lifestyle related suggestions for maintaining a healthy life.
	\item The app offers user rewards like getting smart-watches, exercise equipment and gym memberships.
	\item The app collects your daily statistics such as step count, resting heart rate, geolocation, physical movements.
	\item The collected data will be stored at the app developer side and be shared with the hospitals and clinics.
	\item The collected data will be protected by strict access control.
	\item You can stop data collection and sharing by changing the settings in the app.
\end{itemize}
\vspace{.15in}

Q1. How likely are you to use this app on your smartphone? (On a scale of 1-7, with 1 being very unlikely and 7 being very likely)\\

Q1a. What are the reasons that motivate you to use this app? (Please select all that apply)

\begin{itemize}
	\item I think it may help with my health.
	\item I am interested in the user rewards for using the app.
	\item I am curious to find out more about the app.
	\item  I want to try the app for fun.
	\item I want to share the app experience with my friends.
	\item I think the app is safe to use.
	\item Others. Please specify: 
\end{itemize}

\vspace{.08in}
Q1b. What are the reasons that prevent you from using this app? (Please select all that apply)

\begin{itemize}
	\item I do not think it will help on my health.
	\item The app lacks some important functions.
	\item I am not interested in the user rewards for using the app.
	\item I do not want the app to collect my health data.
	\item I do not want the app to collect my mobile usage data.
	\item I do not want the app to share my health data with other parties.
	\item I do not want the app to share my mobile usage data with other parties.
	\item Using this app may bring potential privacy risks.
	\item Others. Please specify:
\end{itemize}

\vspace{.08in}
Q2. How useful do you rate this app in keeping you healthy? (On a scale of 1-7, with 1 being not useful at all and 7 being very useful)

\vspace{.08in}
Q3. What is your major goal if you decide to use the health app?

\begin{itemize}
	\item Personal health monitoring
	\item Getting personalized feedback from the app
	\item Getting advice from health professionals
	\item Share my health data with my family
	\item It can be used as an assistance data source by my doctors
	\item Getting the user rewards
	\item Out of curiosity
	\item Others. Please specify:
\end{itemize}

\vspace{.08in}
Q4. What do you like most about the health app?

\begin{itemize}
	\item The function of personal health monitoring and disease prevention
	\item The function of personal health data collection
	\item The function of data sharing with other parties
	\item The function of privacy protection
	\item User rewards for using the app
	\item Users' level of control over the app
	\item Others. Please specify:
\end{itemize}

\vspace{.08in}
Q5. What do you like least about the health app?

\begin{itemize}
	\item The function of personal health monitoring and disease prevention
	\item The function of personal health data collection
	\item The function of data sharing with other parties
	\item The function of privacy protection
	\item User rewards for using the app
	\item Users' level of control over the app
	\item Others, please specify
\end{itemize}

\vspace{.15in}

Section II. Open questions
\vspace{.15in}

Q1. Are there any features that you think you need but are missing in the health app? 

\vspace{.08in}
Q2. Describe a situation in which the health app is the most useful to you.

\vspace{.08in}
Q3. Have you noticed any differences among the scenarios? (Please select all that apply)

\begin{itemize}
	\item The types of personal data collected
	\item The feature of user rewards
	\item The feature of data sharing
	\item The feature of app function
	\item Others, please specify
\end{itemize}

\vspace{.15in}
Section III. Background survey

\vspace{.15in}
Part 1/7. Smartphone usage
\vspace{.15in}

Q1. Do you currently have an account with a mobile or cell phone service provider, or not?
\begin{itemize}
	\item Yes, I do
	\item No, I do not
\end{itemize}

\vspace{.08in}
Q2. What is your phone’s operating system?
\begin{itemize}
	\item Android/Google
	\item iOS/Apple
	\item Other (please specify:)
	\item Don’t know
\end{itemize}

\vspace{.08in}
Q3. How many working mobile or cell phones do you currently have?
\begin{itemize}
	\item 1
	\item 2
	\item 3
	\item More than 3
\end{itemize}

\vspace{.08in}
Q4. In a typical weekday, about how much time, in total, do you spend using your mobile or cell phone?
\begin{itemize}
	\item I spend less than one hour on my phone each day
	\item I spend 1 to 3 hours on my phone each day
	\item I spend 4 to 6 hours on my phone each day
	\item I spend 7 to 9 hours on my phone each day
	\item I spend more than 10 hours on my phone each day
\end{itemize}

\vspace{.08in}
Q5. Which of the following activities do you do on your mobile or cell phone? (Please select all that apply)
\begin{itemize}
	\item Browsing websites
	\item Reading and/or writing email
	\item Taking photos/videos
	\item Looking at content on social media websites/apps (for example looking at text, \item images, videos on Facebook, Twitter, Instagram)
	\item Posting content to social media websites/apps (for example posting text, images, videos on Facebook, Twitter, Instagram)
	\item Making purchases (for example buying books or clothes, booking train tickets, ordering food)
	\item Online banking (for example checking account balance, transferring money)
	\item Installing new apps (for example from iTunes, Google Play Store)
	\item Using GPS/location-aware apps (for example Google Maps, Foursquare, Yelp)
	\item Connecting to other electronic devices via Bluetooth (for example smartwatches, fitness bracelets, step counter)
	\item Playing games
	\item Streaming videos or music
	\item Other (please specify:)
\end{itemize}

\vspace{.08in}
Q6. Generally, how would you rate your skills of using your smartphone? (On a scale of 1-7, with 1 being beginner level and 7 being advanced level)

\vspace{.15in}
Part 2/7. Context (user’s health status, COVID-related experience, etc)
\vspace{.15in}

Q1. Are you or have you been infected with Covid-19?
\begin{itemize}
	\item I was tested for Covid-19 and at least one of the results was positive.
	\item I was tested for Covid-19 and all results were negative.
	\item I was not tested for Covid-19, but I suspect that I might have been infected.
	\item I was not tested for Covid-19 and I do not think I have been infected.
\end{itemize}

\vspace{.08in}
Q2. Is there a person in your social circle who is or has been infected with Covid-19?
\begin{itemize}
	\item Yes
	\item No
\end{itemize}

\vspace{.08in}
Q3. Have you been quarantined or did you quarantine yourself because of Covid-19?
\begin{itemize}
	\item Yes
	\item No
\end{itemize}

\vspace{.08in}
Q4. How concerned are you that you or someone you are close to will become infected with Covid-19?
\begin{itemize}
	\item Not at all concerned
	\item Slightly concerned
	\item Somewhat concerned
	\item Moderately concerned
	\item Extremely concerned
\end{itemize}

\vspace{.08in}
Q5. Are you using or have you used any apps for contact tracing?
\begin{itemize}
	\item No, I have never used any of them
	\item I have used such apps but I am not using them anymore
	\item Yes, I am using such apps
\end{itemize}

\vspace{.08in}
Q6. Please tell us your opinion about the contact tracing apps.
\begin{itemize}
	\item Totally unacceptable, they breach my personal privacy
	\item Unacceptable
	\item Netural
	\item Acceptable only under the Covid-19 situation
	\item Acceptable even without Covid-19
\end{itemize}

\vspace{.08in}
Q7. In the past, private companies have shared their customers’ data, such as phone location data, with public authorities to help limit the spread of the Covid-19 pandemic. How do you rate this practice?
\begin{itemize}
	\item Totally unacceptable
	\item Unacceptable
	\item Netural
	\item Acceptable
	\item Perfectly acceptable
\end{itemize}

\vspace{.08in}
Q8. Do you have any Chronic conditions or chronic disease?
\begin{itemize}
	\item Yes
	\item No
\end{itemize}

\vspace{.08in}
Q9. If you choose yes in the above question, please let us know which chronic condition you have (please select all that apply). Otherwise, please leave it blank.
\begin{itemize}
	\item Diabetes
	\item Asthma
	\item High blood pressure
	\item High cholesterol
	\item Arthritis
	\item Alzheimer's disease
	\item Cancer
	\item Others, please specify
\end{itemize}

\vspace{.15in}
Part 3/7. Prior experience with health apps and wearable devices
\vspace{.15in}

Q1. Have you used any health apps?
\begin{itemize}
	\item Yes
	\item No
	\item Not sure
\end{itemize}

\vspace{.08in}
Q1a. If no: Why do you not use a health app?

\vspace{.08in}
Q1b. If yes: please let us know which type of health app you have used (please select all that apply). Otherwise, please leave it blank.
\begin{itemize}
	\item Health app preinstalled in iPhone
	\item Health app preinstalled in Android phones
	\item Health app provided by wearable devices
	\item Health app for recording steps
	\item Health app for monitoring heart rate
	\item Health app for women (e.g., period monitoring, pregnancy, etc)
	\item Health app provided by local health organizations
	\item Health app provided by local government
	\item Others, please specify
\end{itemize}

\vspace{.08in}
Q2. If you have used any health apps, please let us know how likely you would recommend the health app(s) to your family members and friends. Otherwise, please leave it blank.

\begin{itemize}
	\item Very unlikely
	\item Somehow unlikely
	\item Somehow likely
	\item Very likely
\end{itemize}

\vspace{.08in}
Q3. In general, what do you consider positive aspects of using a health app?

\vspace{.08in}
Q4. In general, what do you consider negative aspects of using a health app?

\vspace{.08in}
Q5. Have you used any wearable devices?
\begin{itemize}
	\item Yes
	\item No
\end{itemize}

\vspace{.08in}
Q5a. If no: Why do you not use a wearable device?

\vspace{.08in}
Q5b. If yes: please let us know which type of wearable devices you have used (please select all that apply). Otherwise, please leave it blank.

\begin{itemize}
	\item Apple smart watch
	\item Fitbit Flex
	\item Sumsung smartwatch
	\item Jawbone UP
	\item Xiaomi smart watch
	\item Wearable device for recording steps
	\item Wearable device for monitoring heart rate
	\item Wearable device provided by local health organizations
	\item Wearable device provided by local government
	\item Others, please specify
\end{itemize}

\vspace{.08in}
Q6. If you have used any wearable devices, please let us know which type of data does your wearable devices collect (please select all that apply). Otherwise, please leave it blank.

\begin{itemize}
	\item Step count
	\item Heart rate/pulse rate
	\item Weight
	\item Height
	\item Temperature
	\item Blood pressure
	\item Medication
	\item Medical records
	\item Daily diet
	\item Location
	\item Others, please specify
\end{itemize}

\vspace{.08in}
Q7. If you have used any wearable devices, please let us know how likely you would recommend the wearable device(s) to your family members and friends. Otherwise, please leave it blank.

\begin{itemize}
	\item Very unlikely
	\item Somehow unlikely
	\item Somehow likely
	\item Very likely
\end{itemize}

\vspace{.08in}
Q8. In general, what do you consider positive aspects of using a wearable device?

\vspace{.08in}
Q9. In general, what do you consider negative aspects of using a wearable device?

\vspace{.15in}
Part 4/7. Individual privacy concerns.
\vspace{.15in}

Q1. On a scale of 1 to 5, to what extent do you generally agree with the following statements about your use of the mobile Internet and your privacy? (1: strongly disagree, 5: strongly agree)

\vspace{.08in}
(1) I'm worried about companies having access to my profile	

(2) I'm worried that my information can be more easily accessed by others through a mobile device than by other means		

(3) I'm worried about the privacy of my health records if I were to use mobile health applications		

(4) Sharing my health information on my social network is not a concern	

(5) Privacy issues and my mobile data activities are not a concern	

(6) Making transactions on my mobile phone is not a concern		

(7) I'm more comfortable using my computer for things involving my personal information than using my cell phone

\vspace{.08in}
Q2. How much do you concern about the privacy breach on mobile apps? (On a scale of 1-7, with 1 being no concern at all and 7 being strong concern)

\vspace{.08in}
Q3. Has news about privacy breaches affected your behavior on mobile usage?

\begin{itemize}
	\item No it does not affect my behavior on mobile usage
	\item Yes it slightly affected my behavior on mobile usage
	\item Yes it greatly affected my behavior on mobile usage
	\item Others. Please specify:
\end{itemize}

\vspace{.08in}
Q4. If you choose yes in the above question, please specify how it changed your behavior (please select all that apply). Otherwise please leave it blank.

\begin{itemize}
	\item I download mobile apps less frequently
	\item I use mobile apps less frequently
	\item I think more carefully before downloading an app
	\item I check the privacy settings of mobile apps more carefully
	\item Others. Please specify:
\end{itemize}

\vspace{.15in}
Part 5/7. Privacy Concern (IUIPC scale)
\vspace{.15in}

(1) IUIPC1 For this part of the survey, we are interested in your opinions about managing your privacy when online. There is no right or wrong answer, please use the full 7-point scale in responding, where 1 stands for ``Strongly Disagree'' and 7 for ``Strongly Agree''.

(2) Consumer (user) online privacy is really a matter of consumers’ (users’) right to exercise control and autonomy (independence) over decisions about how their information is collected, used and shared.

(3) Consumer (user) control of personal information lies at the heart of consumer privacy.	

(4) I believe that online privacy is invaded when control is lost or unwillingly reduced as a result of a marketing transaction.	

(5) Companies seeking information online should disclose the way data are collected, processed and used.	

(6) A good consumer online privacy policy should have an easily visible and clear disclosure.	

(7) It is very important to me that I am aware and knowledgeable about how my personal information will be used.

(8) It usually bothers me when online companies ask me for personal information.		

(9) When online companies ask me for personal information, I sometimes think twice about providing it.	

(10) It bothers me to give personal information to so many online companies.		

(11) I’m concerned that online companies are collecting too much personal information about me.

\vspace{.08in}
Q2. How much have you heard or read about the use and potential misuse of information collected from the Internet? (On a scale of 1-7, with 1 being not having heard or read at all and 7 being having heard and read very much)

\vspace{.08in}
Q3. In general, how worried are you about your personal privacy? On a scale of 1-7, with 1 being no concern at all and 7 being strong concern)

\vspace{.08in}
Q4. Please tell us your reason for the above question (why you concern/not concern about personal privacy)

\vspace{.08in}
Q5. I’m protecting my personal privacy so that (please select all that apply)

\begin{itemize}
	\item My life will be safer
	\item My family will be safer
	\item To protect my personal image
	\item To avoid potential financial loss
	\item To avoid potential insurance loss
	\item To have more freedom
	\item To avoid advertisements that I don’t want
	\item To maintain my friendship
	\item To protect my job/business
	\item To avoid social contact with others
	\item Others. Please specify:
\end{itemize}

\vspace{.08in}
Q6. In the past, please rate how many times you provided following information online. (On a scale of 1-5, with 1 being had never provided this information online and 5 being frequently provided this information online)

\begin{itemize}
	
	\item Name	 	
	
	\item Home mailing address	
	
	\item Business mailing address	 
	
	\item Email address	 
	
	\item Social security number	 
	
	\item Date of birth	 
	
	\item Family information (children’s name/ages, marital status)	 
	
	\item Product preferences	 
	
	\item Credit card/banking/stock portfolio information	 
	
	\item Medical information	 
	
	\item Salary/resume information	 
\end{itemize}

\vspace{.15in}
Part 6/7. Attitudes towards changes and high tech
\vspace{.15in}

Q1. Please rate the following statement on a scale of 1-7, with 1 being strongly disagree and 7 being strongly agree.

(1) Technology is my friend.

(2) I relate well to technology and machines.	

(3) I am comfortable learning new technology.	

(4) I know how to deal with technological malfunctions or problems.	

(5) I find most technology easy to learn.			

(6) I feel as up-to-date on technology as my peers.

\vspace{.08in}
Q2. If ‘technophobia’ is defined as feeling discomfort about computers or any new technology, which of the following best describes you:

\begin{itemize}
	\item Not Technophobic
	\item Mildly Technophobic
	\item Moderately Technophobic
	\item Highly Technophobic
\end{itemize}

\vspace{.15in}
Part 7/7. Basic demographics
\vspace{.15in}

Q1. Your gender:
\begin{itemize}
	\item Male
	\item Female
	\item Others
\end{itemize}

\vspace{.08in}
Q2. Your race:
\begin{itemize}
	\item American Indian or Native American
	\item Asian / Pacific Islander
	\item Black or African American
	\item Hispanic or Latino
	\item White
	\item Other
\end{itemize}

\vspace{.08in}
Q3. Your age:
\begin{itemize}
	\item $<$19
	\item 19-24
	\item 25-34
	\item 35-44
	\item 45-59
	\item 60+
\end{itemize}

\vspace{.08in}
Q4. Your education:
\begin{itemize}
	\item Below high school
	\item High school graduate (includes equivalency)
	\item Trade/technical/vocational training
	\item Bachelor's degree
	\item Graduate degree (e.g., M.S. and Ph.D.)
\end{itemize}

\vspace{.08in}
Q5. In what country do you currently live?
\begin{itemize}
	\item United States
	\item India
	\item China
	\item Singapore
	\item Canada
	\item United Kingdom
	\item Germany
	\item Other European countries
	\item Other Asian countries
	\item Australia
	\item New Zealand
	\item None of the above
\end{itemize}

\end{document}